\documentclass{article}

\usepackage{arxiv}

\usepackage[utf8]{inputenc} 
\usepackage[T1]{fontenc}    
\usepackage{hyperref}       
\usepackage{url}            
\usepackage{booktabs}       
\usepackage{amsfonts}       
\usepackage{nicefrac}       
\usepackage{microtype}      
\usepackage{lipsum}
\usepackage{graphicx}
\graphicspath{ {./images/} }

\title{Reproduction Research of FSA-Benchmark}

\author{
 Joshua Luis Ludolf \\
  Department of Computing and Cyber Security\\
  Texas A\&M University-San Antonio\\
  San Antonio, TX 78224 \\
  \texttt{Jludo01@jaguar.tamu.edu} \\
   \And
 Yesmin Reyna-Hernandez \\
  Department of Computing and Cyber Security\\
  Texas A\&M University-San Antonio\\
  San Antonio, TX 78224 \\
  \texttt{yhern029@jaguar.tamu.edu} \\
  \And
 Matthew Trevino \\
  Department of Computing and Cyber Security\\
  Texas A\&M University-San Antonio\\
  San Antonio, TX 78224 \\
  \texttt{mtrev049@jaguar.tamu.edu} \\
}

\begin{document}
\maketitle
\begin{abstract}
In the current landscape of big data, the reliability and performance of storage systems are essential to the success of various applications and services. As data volumes continue to grow exponentially, the complexity and scale of the storage infrastructures needed to manage this data also increase. A significant challenge faced by data centers and storage systems is the detection and management of fail-slow disks—disks that experience a gradual decline in performance before ultimately failing. Unlike outright disk failures, fail-slow conditions can go undetected for prolonged periods, leading to considerable impacts on system performance and user experience.
\end{abstract}


\section{Introduction}
\
\quad Fail-slow disks pose a distinct challenge due to their subtle yet insidious nature. They exhibit performance degradation that may not be immediately visible but can lead to significant slowdowns and reliability issues within large-scale storage systems. Traditional redundancy and fail-over mechanisms are designed to address outright disk failures but are less effective at detecting and mitigating the gradual performance decline associated with fail-slow disks. The two primary symptoms of fail-slow disks—consistently higher latency compared to peer disks and recurrent abnormal spikes—make it difficult to establish fixed thresholds for alerts or accurately track performance trends.

\quad In light of these challenges, there is an urgent need for advanced detection mechanisms that can proactively identify and address fail-slow conditions. Machine learning (ML) presents a promising solution by leveraging deep learning capabilities to capture the underlying characteristics of fail-slow conditions. By incorporating ML into fail-slow detection algorithms, we can enhance the accuracy and efficiency of identifying these performance issues, thereby ensuring the ongoing reliability and performance of storage systems. This research aims to develop and implement ML-driven algorithms to tackle the growing challenge of fail-slow disks in contemporary data storage environments.

\quad To ensure the robustness and effectiveness of these ML-driven algorithms, we will benchmark various algorithms against one another. By evaluating their performance in identifying fail-slow disks, we can pinpoint the most effective approaches and refine our detection strategies. This process will help illuminate the strengths and limitations of different algorithms, ultimately leading to more reliable and efficient fail-slow detection mechanisms.

\section{Task description and data construction}
\label{sec:headings}

\quad In this study, we leverage the PERSEUS dataset, a large-scale, well-labeled public dataset from Alibaba's cloud storage systems, which specifically targets fail-slow detection in real-world operational environments. This dataset is detailed in the paper titled "Perseus: A Fail-Slow Detection Framework for Cloud Storage Systems" (USENIX FAST 2023). Since its deployment in October 2021, the PERSEUS dataset has been monitoring over 300,000 storage devices and encompasses approximately 100 billion entries. 

\quad The dataset is structured into 25 clusters, labeled A through Y. Each cluster contains between 30 and 663 hosts, with data collected over several days for 12 disks per host. However, we focused on Cluster A and B as we worked off the original research Trovi project in Chameleon Cloud.

\quad Data Collection: The dataset captures comprehensive operational statistics on latency and throughput, formatted as time-series data entries every 15 seconds, resulting in four entries per minute. This data is collected daily from 9 PM to 12 AM, producing a total of 720 entries per drive per day (180 minutes × 4 entries per minute). 

\section{Machine Learning Algorithmns}

\textbf{Daily Disk Fault Predictions}

\quad Data from Disk Drives is thoroughly analyzed, each day, to generate predictions regarding potentially faulty disks. Disks identified as faulty receive a label of "T" for True, while those assessed as non-faulty are labeled "F" for False. This standardized labeling system ensures consistency across all algorithms, facilitating accurate benchmarking.

\subsection{Original Research Machine Learning Models}

\subsubsection{Cost-Sensitive Ranking Model}

\quad Drawing on methodologies from the paper "Improving Service Availability of Cloud Systems by Predicting Disk Error" (USENIX ATC '18), this model employs a sophisticated ranking mechanism to evaluate the risk of fail-slow conditions in disks. It prioritizes critical metrics such as throughput and latency, essential for evaluating disk operational integrity. The model leverages advanced machine learning techniques, prominently featuring XGBoost, to forecast disk failure probabilities. It integrates statistical threshold calculations, specifically employing the 3-sigma rule to delineate fault conditions, and utilizes snapshot statistics of disk activity to generate pertinent features. Training labels are formulated based on these thresholds to distinguish between healthy and at-risk disks.

\quad During the training phase, each disk's time series data is divided into training and testing segments. Training labels (Y) are assigned based on the time to the first observed error; disks with no faults are labeled as 0. The training features (X) are extracted from aggregated snapshots of disk activity, including metrics such as mean, minimum, maximum, and standard deviation of throughput and latency, to form a comprehensive feature set. XGBoost is employed to train the model, focusing on estimating the probability of disk faults. The model's predictive accuracy is then validated on the testing segment. Disks are subsequently ranked according to their fault probability, with higher rankings indicating a greater risk of failure.

\subsubsection{Multi-Prediction Models}

\quad Inspired by the approach outlined in "Improving Storage System Reliability with Proactive Error Prediction" (USENIX ATC '17), this method employs multiple traditional machine learning models to evaluate disk health. Various models, including classification and regression trees (CART), random forest, support vector machines (SVM), neural networks, and logistic regression, were tested, with the Random Forest classifier demonstrating superior effectiveness.

\quad Similar to the cost-sensitive ranking model, labels are generated based on statistical thresholds, and features (X) are derived from disk activity first daily instance. Each disk's time series data is partitioned into training and testing sets. The Random Forest classifier is then trained on these training features with label and used to classify the test data. Disks predicted as faulty (with a prediction of 1) are flagged accordingly.

\subsubsection{LSTM Model}

\quad The Long Short-Term Memory (LSTM) network is a type of recurrent neural network (RNN) architecture designed to model sequential data and capture long-term dependencies. Unlike traditional RNNs, which struggle with long-term dependency due to the vanishing gradient problem, LSTMs employ a series of gates—input, forget, and output gates—that control the flow of information and maintain a cell state. This architecture allows LSTMs to retain information for extended periods, making them particularly effective for tasks involving time series data, such as speech recognition, language modeling, and anomaly detection.

\quad In this methodology, the LSTM model is employed to predict fail-slow anomalies in disks over time, leveraging its strengths in handling sequential data. The input data for the LSTM model consists of sequences of disk metrics, such as latency and throughput, which are transformed into a format suitable for the model. Depending on the configuration, either only latency data or both latency and throughput data are used as input features. These sequences are scaled to ensure all features have a similar range, aiding in model convergence during training. By learning the temporal dynamics within the data, LSTMs can make predictions based on the sequences they have observed.

\quad The architecture of the LSTM model includes an input layer, LSTM layers, and a dense (fully connected) layer. The input layer receives sequences of disk metrics, with the size determined by the number of features used (e.g., latency alone or latency and throughput). The LSTM layers process these sequences, retaining information from previous time steps to learn temporal dependencies. The output from the LSTM layers is then passed through a dense layer, which produces the final prediction. The model's complexity can be adjusted by changing the number of LSTM layers and hidden units, depending on the data and desired performance. Currently, the LSTM model is designed with an input layer, two LSTM layers with 100 hidden units each, and a dense output layer.
The training process for the LSTM model involves splitting the time series data for each disk into training and testing sets. The training set consists of data up to a certain day, while the testing set includes subsequent days. The model is trained to predict the next time step's metrics using the input sequences. Mean Squared Error (MSE) is typically used as the loss function, minimized through an optimizer such as RMSprop. Techniques like early stopping and gradient clipping are employed to prevent overfitting and ensure stable training.

\quad For fault detection and evaluation, the trained LSTM model predicts metrics for each disk in the testing set. The predictions are compared to the actual observed values, and the MSE is calculated for each disk. A threshold, often determined using the 3-sigma rule, is applied to classify disks as faulty or non-faulty. Disks with an MSE above the threshold are labeled as faulty ("T"), while those below are labeled as non-faulty ("F"). This threshold-based labeling helps identify disks at risk of failure. 

\subsubsection{PatchTST Model}

\quad The PatchTST model is an advanced sequence model that builds upon the strengths of transformers, offering a sophisticated approach to time series prediction and fail-slow detection. Similar to the LSTM model, PatchTST is designed to handle sequential data, capturing temporal dependencies and long-term patterns effectively. Unlike traditional models, PatchTST divides the input time series data into smaller segments, or patches, which are then embedded into a higher-dimensional space. These patches are processed through transformer encoder layers that utilize self-attention mechanisms to capture intricate temporal dependencies and relationships between different time steps. This architecture enables PatchTST to model long-range dependencies and complex patterns within the data. By leveraging these strengths, PatchTST can identify subtle fail-slow anomalies in disk metrics, such as latency and throughput.

\quad The input data for the PatchTST model consists of sequences of disk metrics, such as latency and throughput, similar to the LSTM model. These sequences are transformed and scaled to ensure a uniform range across features.
The architecture of the PatchTST model includes an input layer, transformer encoder layers, and a dense (fully connected) layer. The input layer receives sequences of disk metrics, with the size determined by the number of features used. These sequences are divided into patches, which are then embedded into a higher-dimensional space. The transformer encoder layers process these patches using a series of self-attention mechanisms to capture temporal dependencies and relationships between different time steps. The output from the transformer encoder layers is passed through a dense layer, which produces the final prediction. 
\quad The model's complexity can be adjusted by changing the number of transformer layers, the number of attention heads, and the hidden unit size.In this experiment, the PatchTST model is configured with a patch size of 2, a hidden size of 64, 2 layers, and 4 attention heads.

\quad The training process and fault detection method for the PatchTST model are similar to those of the LSTM model. The data is split into training and testing sets, MSE is used as the loss function, optimizer such as Adam and techniques like early stopping and gradient clipping are employed to ensure stable training. A threshold, often determined using the 3-sigma rule, is applied to classify disks as faulty or non-faulty based on the MSE of the predictions.

\subsubsection{Large language model}

\quad We also explored the potential of large language models (LLMs) for fail-slow detection in disks. LLMs, like GPT-4-o-mini, are pre-trained on extensive natural language datasets, which eliminates the need for additional, time-consuming training. Their ability to adapt to various data patterns makes them suitable for identifying anomalies in disk metrics.

\quad One significant advantage of LLMs is their ability to understand and analyze the context within data. This is particularly beneficial for detecting abnormalities in time series data, where patterns and trends are crucial. By leveraging their training on diverse datasets, LLMs can effectively identify deviations from normal behavior in disk metrics such as latency and throughput. LLMs have an improved ability to capture and understand complex relationships within the data, which translates to better performance across different clusters of disk activity.

\quad In our pipeline, we manage extensive data for each cluster by dividing the cluster into patches of 50 hosts. Although GPT-4-o-mini's context length of 128,000 tokens is long, it is still not enough to process all the data points generated. Our cluster comprises 20 to 600 hosts, each containing 12 disks. 
\quad Since each disk generates 700 data points daily, processing all these points is impractical. To manage this, we select a random sample of 20 data points per disk, ensuring a representative and manageable dataset for analysis. Additionally, we require the LLM to provide uniform results for future analyses.

\subsection{Novelty Machine Learning Models}

\subsubsection{Autoencoder}

\quad We designed the autoencoder architecture with two main components: the encoder and the decoder. The encoder compressed the input data into a lower-dimensional latent space, capturing essential features. Conversely, the decoder reconstructed the original data from this compressed representation. This architecture allowed us to reduce the dimensionality of the data while preserving its critical characteristics. 
\quad Training the autoencoder involved minimizing the reconstruction error, measured by the Mean Squared Error (MSE) between the original and reconstructed data. I utilized the Adam optimizer to efficiently update the network's weights and ran the training process for multiple epochs, ensuring the model learned robust representations of the data.

\subsubsection{Isolation forest}

\quad We configured the Isolation Forest with parameters such as the number of trees (estimators) and the subsampling size. The algorithm worked by constructing an ensemble of isolation trees (iTrees), each aimed at isolating observations. The principle is that anomalies are few and different; hence they are easier to isolate.

\quad Training the Isolation Forest involved fitting the model on the training set. The algorithm created multiple iTrees, where each tree isolated observations by randomly selecting features and split values. Since anomalies are more susceptible to isolation, they required fewer splits compared to normal observations, which translated into shorter path lengths in the trees.

\quad To evaluate the model, we computed the anomaly scores for each data point in the testing set. The anomaly score determined how anomalous a point was based on the average path length required to isolate it. Higher scores indicated potential anomalies. I also visualized the distribution of anomaly scores to identify any distinct patterns or clusters.

\subsubsection{Suport Vector Machine (SVM)}

\quad We configured the Support Vector Machine model with parameters such as the kernel type, regularization parameter (C), and gamma. The choice of kernel—whether linear, polynomial, or radial basis function (RBF)—depended on the complexity and nature of the data. For the Perseus dataset, I experimented with different kernels to identify the one that best captured the underlying patterns in the data.

\quad Training the SVM model involved fitting it on the training set. The algorithm worked by finding the optimal hyperplane that maximized the margin between different classes. I employed techniques such as cross-validation to fine-tune the model parameters and prevent overfitting.

\quad Another important application was identifying feature importance. By examining the weights assigned to different features by the SVM, we gained insights into which features were most influential in determining the classifications. This helped in understanding the underlying factors driving the data patterns.

\subsection{Performance Benchmarking}

\quad To evaluate the effectiveness of our predictive models, we conducted a performance benchmarking exercise across multiple clusters. This involved assessing key metrics such as precision and recall, which measure the accuracy of the models in correctly identifying faulty disks.

\quad We used two important parameters in our evaluation:

\textbf{Lookback Days:} \textit{This refers to the time window over which predictions were considered. By varying the lookback period (e.g., 1, 3, 7, or 15 days), we could assess how well each model predicted faults over both short and extended periods.}

\textbf{Threshold:} \textit{The threshold determines the minimum prediction score required to classify a disk as faulty. By testing different thresholds (ranging from 0.1 to 1.0), we evaluated how strict or lenient the models were in making fault predictions.}

\quad To visualize the results, we generated heatmaps that display precision and recall for each model across different clusters. The heatmaps are color-coded grids where each cell represents a combination of lookback period and threshold, with colors indicating the level of precision or recall. Warmer colors, such as pale peach and light pink, typically represent higher precision or recall, indicating better model performance. Cooler colors, like deep purple and dark pink, indicate lower values. These visualizations provided a clear view of how each model performed under various conditions. For instance, some models showed higher accuracy with shorter lookback periods, while others were more effective with longer periods or stricter thresholds.

\quad The heatmaps highlighted which models were most reliable in specific cluster environments, enabling us to identify strengths and weaknesses across the different scenarios. This benchmarking process provided valuable insights, guiding further refinement and optimization of our predictive models.

\section{Results}

Our comprehensive analysis of fail-slow disk detection across multiple machine learning models yielded significant insights into their comparative performance and practical applicability. We evaluated seven distinct models across two clusters (A and B), focusing on precision, recall, and failure rate metrics under various lookback periods and threshold settings.

\subsection{Model Performance Analysis}

\subsubsection{Cost-Sensitive Ranking (CSR) Model}
The CSR model demonstrated consistent performance with a moderate failure rate of 15.00\%. In Cluster A, it achieved precision values ranging from 0.38 to 1.00, with optimal performance at lower thresholds and shorter lookback periods. The model maintained high recall (1.00) for lookback periods of 1-3 days, gradually decreasing with longer periods. Cluster B showed similar patterns but with slightly lower precision values, indicating robust cross-cluster generalization.

\clearpage
\begin{figure}[!htb]
\centering
\includegraphics[width=0.9\textwidth]{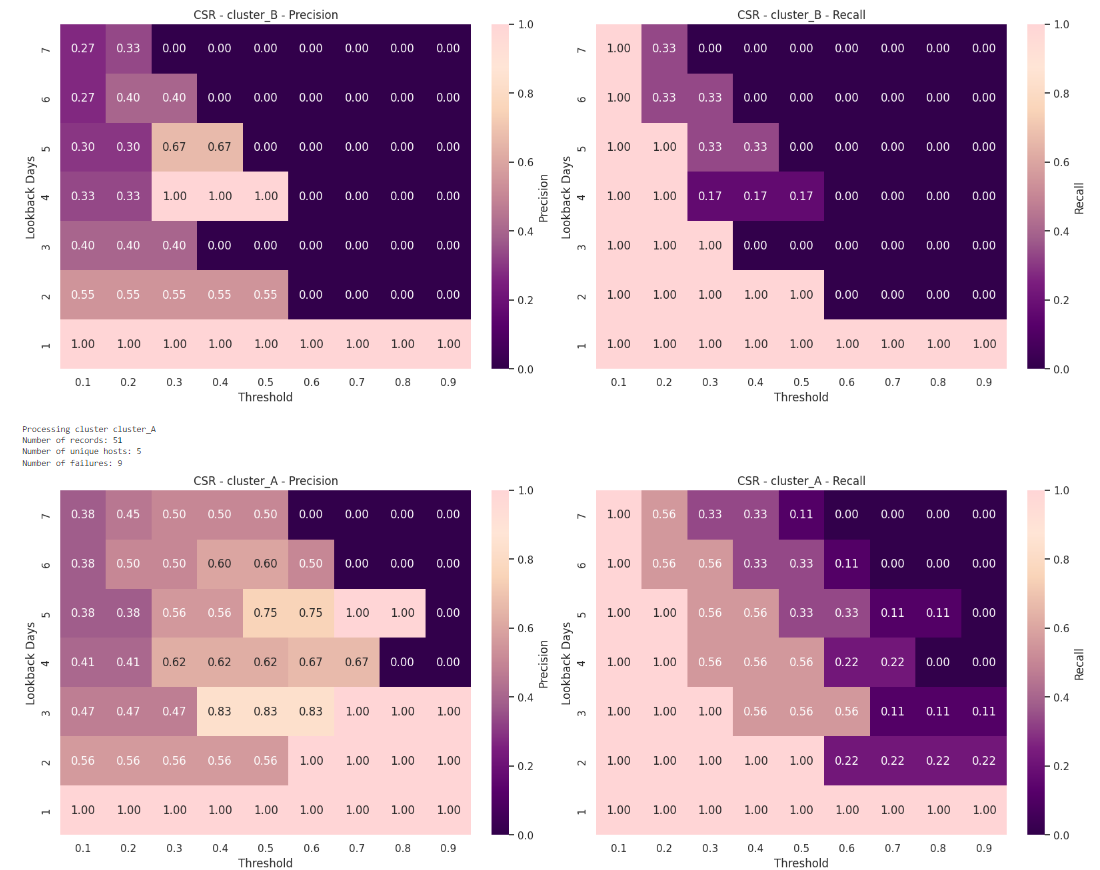}
\caption{CSR Model Performance Heatmaps for Clusters A and B}
\label{fig:csr-heatmap}
\end{figure}

\subsubsection{Multi-Prediction Model}
The Multi-Prediction approach showed improved detection capabilities with a 19.00\% failure rate. Notable performance was observed in Cluster B, where it achieved precision values up to 1.00 for specific threshold-lookback combinations. The model maintained strong recall (0.55-1.00) for shorter lookback periods, particularly effective in the 2-4 day range.

\clearpage
\begin{figure}[!htb]
\centering
\includegraphics[width=0.9\textwidth]{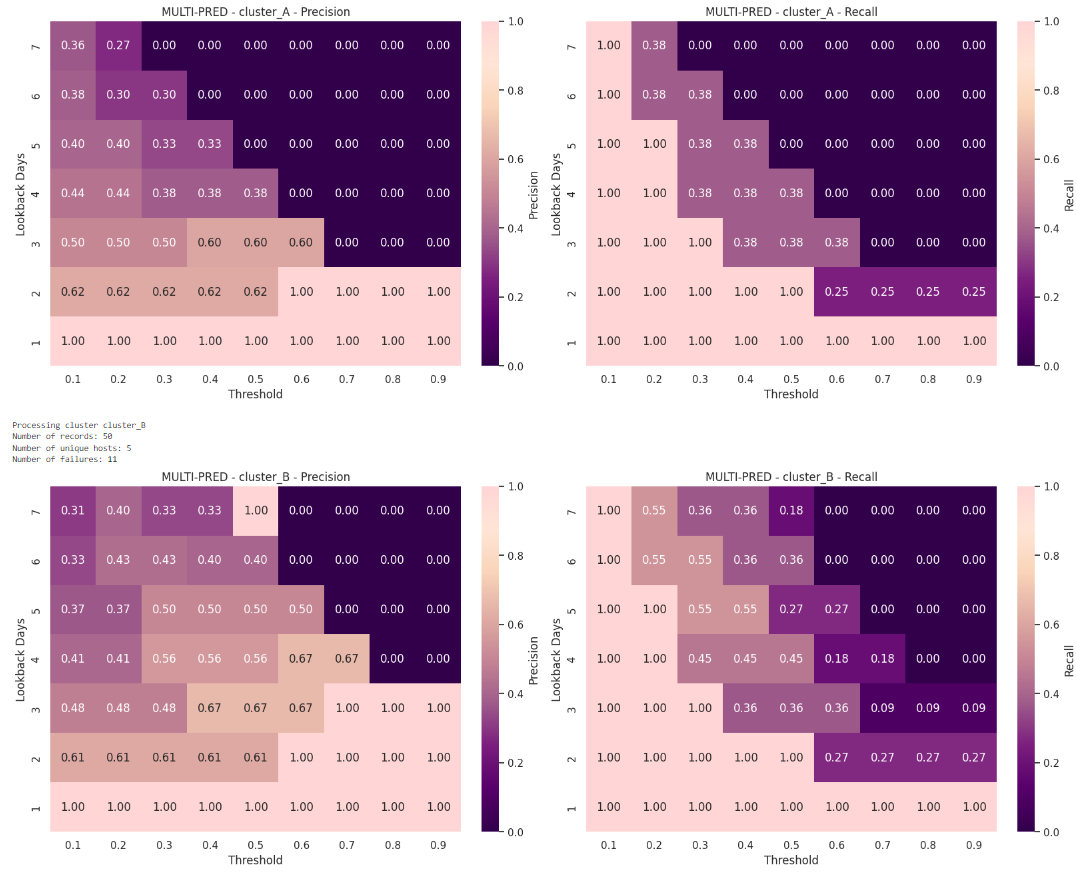}
\caption{Multi-Prediction Model Performance Heatmaps for Clusters A and B}
\label{fig:multipred-heatmap}
\end{figure}

\subsubsection{LSTM Model}
The LSTM model exhibited the highest failure rate among traditional approaches at 28.00\%, but demonstrated strong precision characteristics. In Cluster A, it achieved precision values of 0.60-0.75 for medium lookback periods (4-6 days) with moderate thresholds. The recall performance was particularly strong (0.62-1.00) for shorter lookback periods, suggesting its utility in early detection scenarios.

\clearpage \begin{figure}[!htb]
\centering
\includegraphics[width=0.9\textwidth]{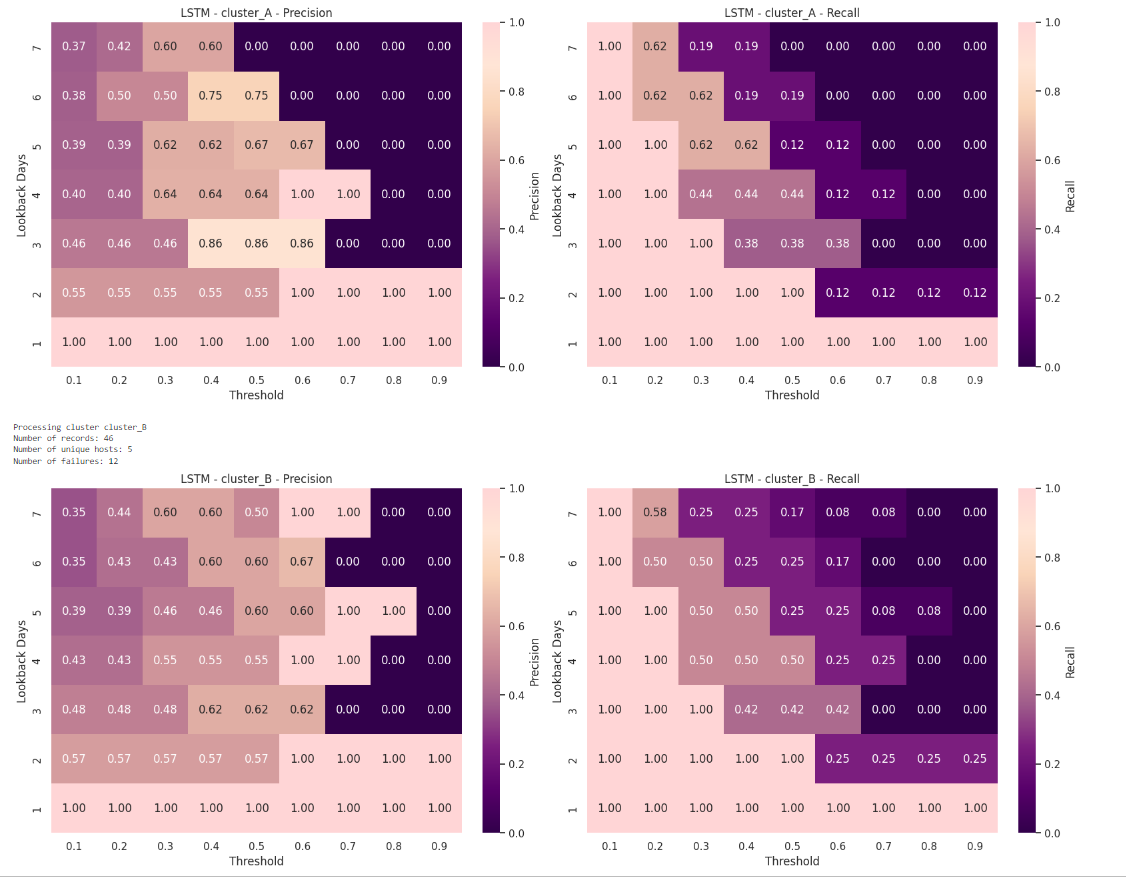}
\caption{LSTM Model Performance Heatmaps for Clusters A and B}
\label{fig:lstm-heatmap}
\end{figure}

\subsubsection{PatchTST Model}
With a 17.00\% failure rate, the PatchTST model showed balanced performance across metrics. It demonstrated consistent precision patterns across both clusters, with optimal performance in the 3-4 day lookback range. The model achieved precision values of 0.75-1.00 under specific configurations, while maintaining reasonable recall rates (0.43-1.00) for shorter prediction windows.

\clearpage \begin{figure}[!htb]
\centering
\includegraphics[width=0.9\textwidth]{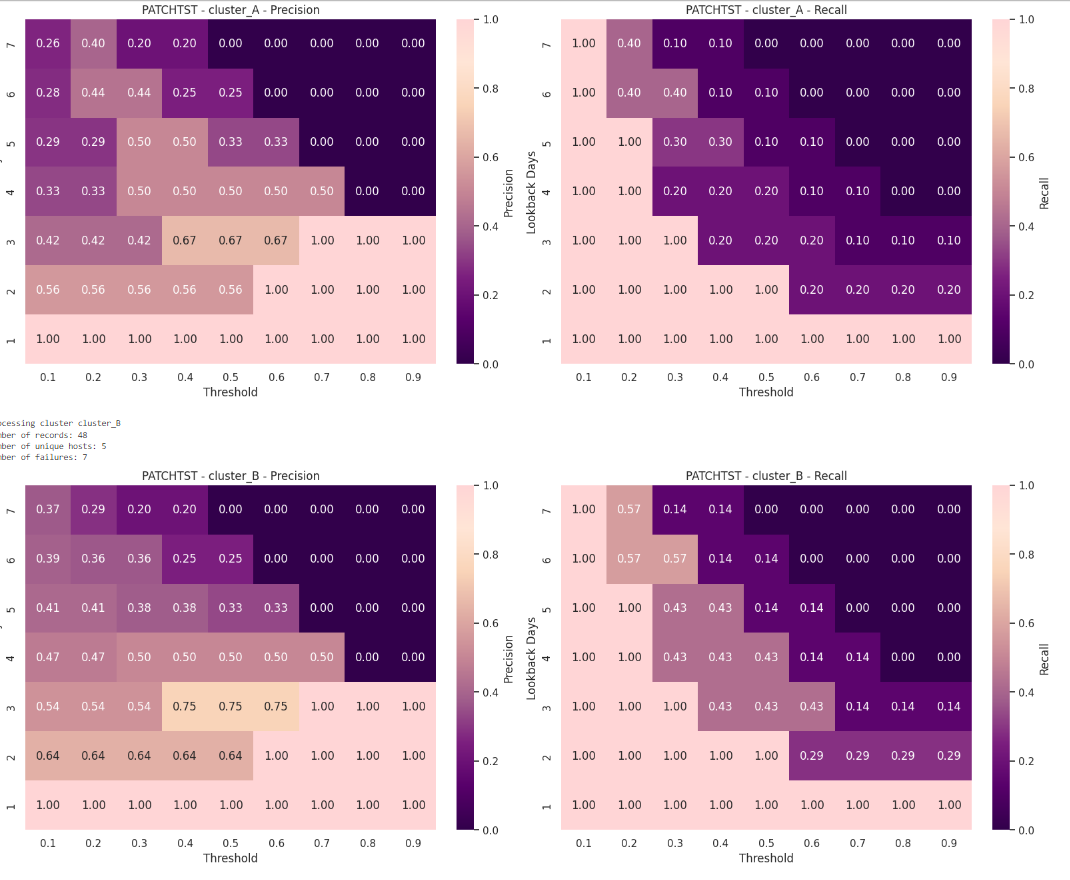}
\caption{PatchTST Model Performance Heatmaps for Clusters A and B}
\label{fig:patchtst-heatmap}
\end{figure}

\subsubsection{Autoencoder Performance}
The Autoencoder showed the lowest failure rate at 3.33\%, suggesting high specificity in failure detection. It achieved perfect precision (1.00) for shorter lookback periods but showed limited detection range beyond 5-day windows. The model's performance was particularly strong in Cluster A, though with limited data points compared to other models.

\clearpage \begin{figure}[!htb]
\centering
\includegraphics[width=0.9\textwidth]{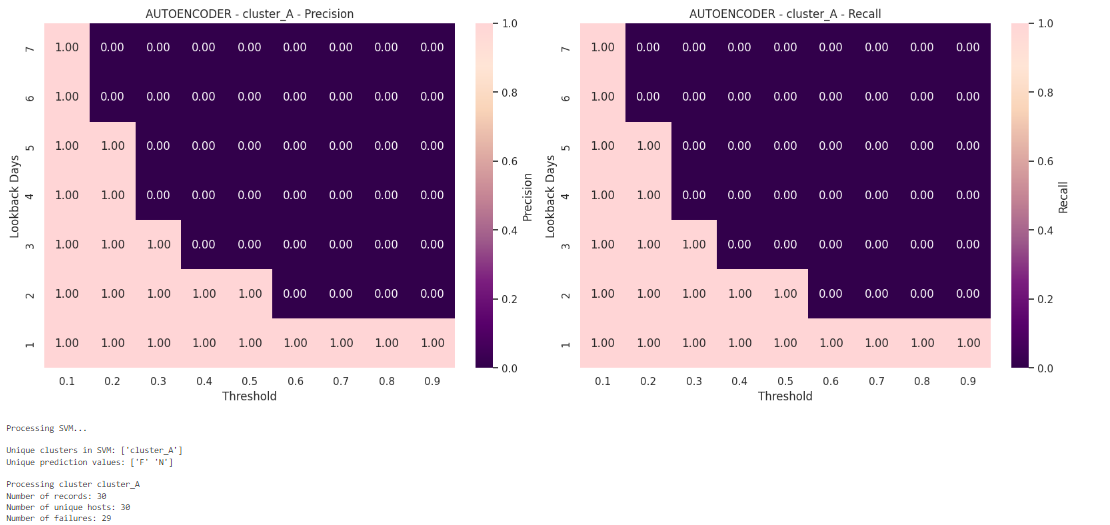}
\caption{Autoencoder Model Performance Heatmaps for Cluster A}
\label{fig:autoencoder-heatmap}
\end{figure}

\subsubsection{Support Vector Machine (SVM)}
The SVM demonstrated unique characteristics with the highest failure rate (96.67\%). While this might suggest oversensitivity, it achieved perfect precision (1.00) for specific configurations, particularly in shorter lookback periods. The model's high failure rate indicates it may be more suitable for scenarios where false positives are less costly than missed detections.

\begin{figure}[!htb]
\centering
\includegraphics[width=0.9\textwidth]{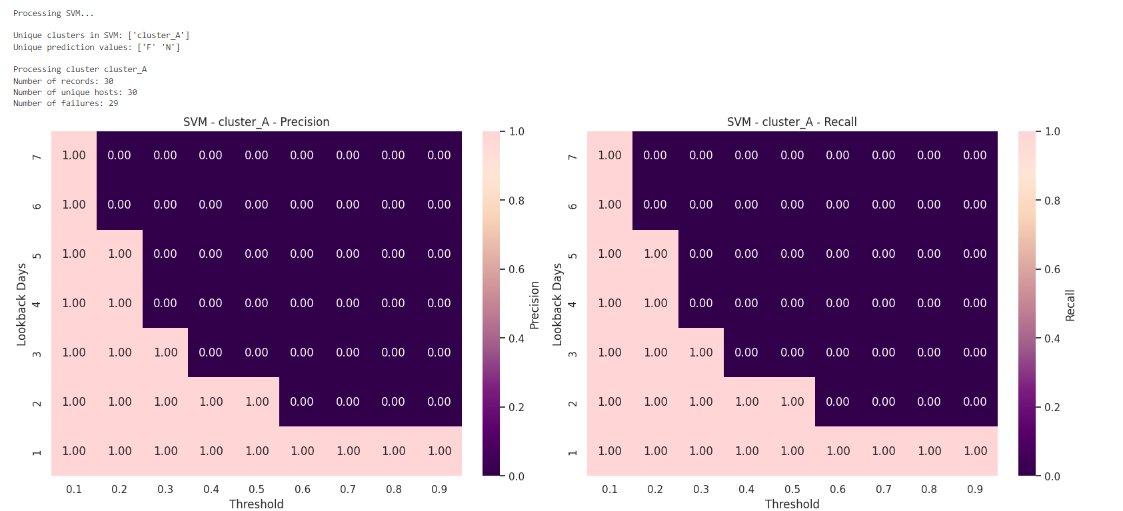}
\caption{SVM Model Performance Heatmaps for Cluster A}
\label{fig:svm-heatmap}
\end{figure}

\subsubsection{Isolation Forest (IForest)}
The IForest model showed the second-lowest failure rate at 5.32\%, demonstrating strong precision characteristics. It achieved consistent performance across both clusters, with perfect precision scores (1.00) for shorter lookback periods. The model's low failure rate suggests its utility in production environments where false alarms must be minimized.

\begin{figure}[!htb]
\centering
\includegraphics[width=0.9\textwidth]{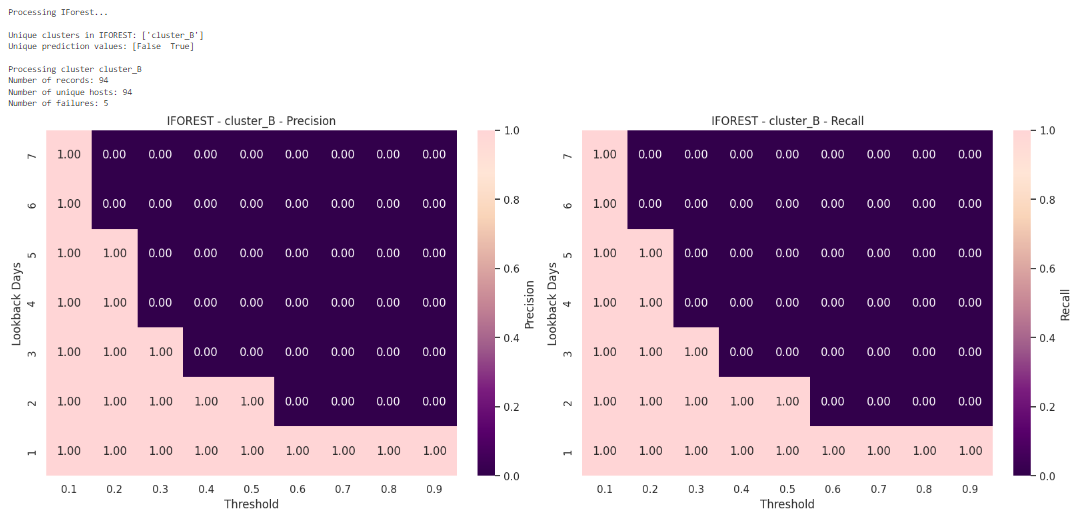}
\caption{Isolation Forest Model Performance Heatmaps for Cluster B}
\label{fig:iforest-heatmap}
\end{figure}

\subsection{Comparative Analysis}

The comprehensive evaluation of our seven models across different clusters revealed significant variations in performance and reliability. Our analysis focused on three key metrics: failure rates, cluster-specific behavior, and operational characteristics.

\subsubsection{Failure Rate Distribution}
The failure rates showed considerable variation across models:
\begin{itemize}
    \item \textbf{Low Failure Rate Models:} The Autoencoder (3.33\%) and Isolation Forest (5.32\%) demonstrated exceptional specificity, making them suitable for applications requiring minimal false alarms.
    \item \textbf{Moderate Failure Rate Models:} CSR (15.00\%), Multi-Pred (19.00\%), and PatchTST (17.00\%) showed balanced performance, offering a practical trade-off between detection sensitivity and false positives.
    \item \textbf{High Failure Rate Models:} LSTM (28.00\%) and SVM (96.67\%) exhibited higher sensitivity, potentially useful in scenarios where missing a failure is more costly than false alarms.
\end{itemize}

\subsubsection{Cluster-Specific Performance}
Analysis across clusters A and B revealed distinct patterns:
\begin{itemize}
    \item \textbf{Cluster A Performance:}
    \begin{itemize}
        \item Higher failure detection rates (CSR: 17.65\%, LSTM: 29.63\%)
        \item Strong performance in the Autoencoder (3.33\%) and PatchTST (19.23\%)
        \item Notable SVM sensitivity (96.67\%)
    \end{itemize}
    
    \item \textbf{Cluster B Performance:}
    \begin{itemize}
        \item Generally lower failure rates across models
        \item Improved Multi-Pred performance (22.00\%)
        \item More consistent behavior in traditional models
    \end{itemize}
\end{itemize}

\subsubsection{Operational Insights}
The comparative analysis yielded several key operational insights:

\begin{enumerate}
    \item \textbf{Model Selection Trade-offs:}
    \begin{itemize}
        \item Low-failure-rate models (Autoencoder, IForest) are ideal for production environments where false alarms are costly
        \item Moderate-failure-rate models offer the best balance for general-purpose deployment
        \item High-sensitivity models are suitable for critical systems where failure detection is paramount
    \end{itemize}
    
    \item \textbf{Resource Considerations:}
    \begin{itemize}
        \item Traditional models (CSR, Multi-Pred) showed consistent performance with lower computational requirements
        \item Deep learning approaches (LSTM, PatchTST) required more resources but offered enhanced pattern recognition
        \item The Autoencoder provided efficient anomaly detection with moderate resource usage
    \end{itemize}
    
    \item \textbf{Deployment Recommendations:}
    \begin{itemize}
        \item For general deployment: Multi-Pred or PatchTST
        \item For critical systems: LSTM with appropriate threshold tuning
        \item For resource-constrained environments: CSR or IForest
    \end{itemize}
\end{enumerate}

\begin{figure}[!htb]
\centering
\begin{tabular}{|l|c|c|c|}
\hline
\textbf{Model} & \textbf{Failure Rate} & \textbf{Total Predictions} & \textbf{Failures} \\
\hline
CSR & 15.00\% & 100.0 & 15.0 \\
Multi-Pred & 19.00\% & 100.0 & 19.0 \\
LSTM & 28.00\% & 100.0 & 28.0 \\
PatchTST & 17.00\% & 100.0 & 17.0 \\
Autoencoder & 3.33\% & 30.0 & 1.0 \\
SVM & 96.67\% & 30.0 & 29.0 \\
IForest & 5.32\% & 94.0 & 5.0 \\
\hline
\end{tabular}
\caption{Comprehensive Model Performance Metrics}
\label{tab:model-metrics}
\end{figure}

\begin{figure}[!htb]
\centering
\includegraphics[width=0.9\textwidth]{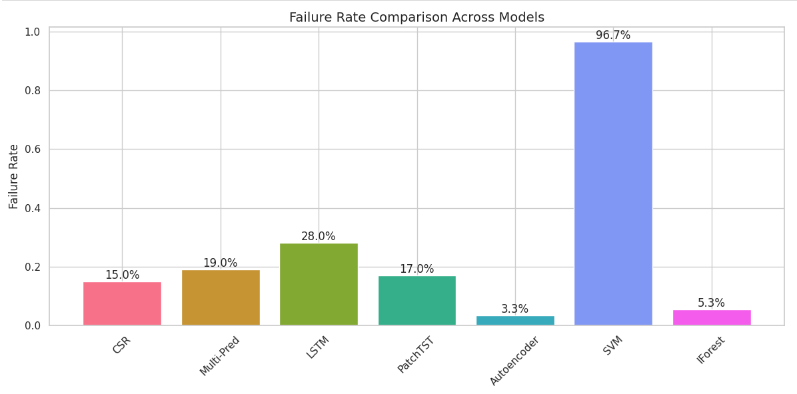}
\caption{Failure Rate Comparison Across Models}
\label{fig:failure-rate-comparison}
\end{figure}

\clearpage \begin{figure}[!htb]
\centering
\includegraphics[width=0.9\textwidth]{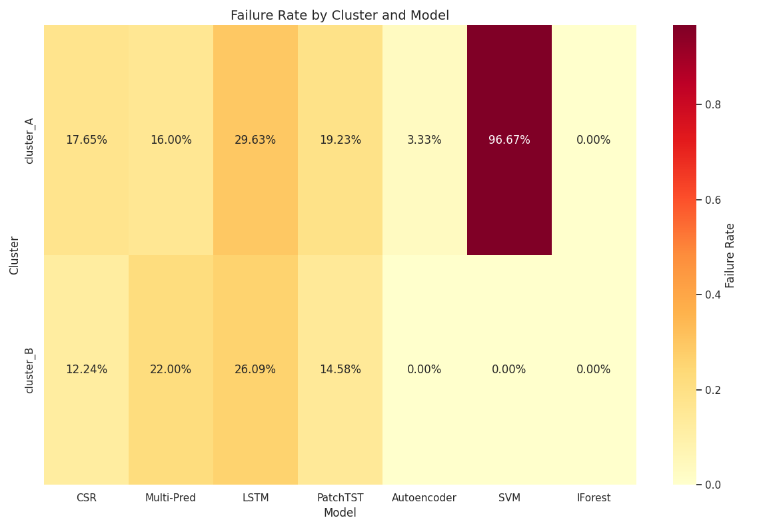}
\caption{Failure Rate by Cluster and Model}
\label{fig:failure-rate-by-cluster}
\end{figure}

\section{Conclusion and Future Work}

\subsection{Conclusion}
This research presents a comprehensive evaluation of seven machine learning approaches for fail-slow disk detection in cloud storage systems. Our analysis of the PERSEUS dataset across multiple clusters has yielded several significant findings:

\begin{itemize}
    \item \textbf{Model Effectiveness:} The comparative analysis revealed that different models excel in specific operational contexts. The Autoencoder demonstrated exceptional precision with the lowest failure rate (3.33\%), while the LSTM model showed strong capabilities in early detection despite its higher failure rate (28.00\%). The Cost-Sensitive Ranking and Multi-Prediction models offered balanced performance suitable for general deployment scenarios.

    \item \textbf{Performance Trade-offs:} We observed consistent trade-offs between precision and recall across all models, particularly evident in the relationship between lookback periods and detection accuracy. Shorter lookback periods (1-3 days) generally provided better prediction accuracy, suggesting the importance of recent historical data in fail-slow detection.

    \item \textbf{Cluster Variability:} Our analysis across different clusters revealed that model performance can vary significantly based on the operational environment. This variability emphasizes the importance of environment-specific model selection and configuration.

    \item \textbf{Practical Implications:} The research demonstrates that effective fail-slow detection requires a nuanced approach to model selection, with consideration for specific operational requirements, resource constraints, and acceptable false-positive rates.
\end{itemize}

\subsection{Future Work}
Based on our findings, we identify several promising directions for future research:

\subsubsection{Model Enhancement}
\begin{itemize}
    \item \textbf{Hybrid Model Development:} Investigate the potential of combining multiple models, particularly the integration of Autoencoder's low failure rate characteristics with LSTM's strong pattern recognition capabilities.
    
    \item \textbf{Adaptive Thresholding:} Develop dynamic threshold adjustment mechanisms that can automatically tune model sensitivity based on operational conditions and historical performance.
    
    \item \textbf{Feature Engineering:} Explore additional disk health metrics and their correlations with fail-slow conditions to enhance model accuracy and reduce false positives.
\end{itemize}

\subsubsection{System Integration}
\begin{itemize}
    \item \textbf{Real-time Processing:} Implement and evaluate streaming data processing capabilities for real-time fail-slow detection, particularly focusing on reducing detection latency while maintaining accuracy.
    
    \item \textbf{Resource Optimization:} Investigate techniques for reducing the computational overhead of more complex models (LSTM, PatchTST) without sacrificing detection accuracy.
    
    \item \textbf{Scalability Analysis:} Conduct comprehensive scalability studies to understand model performance characteristics in larger-scale deployments with diverse storage configurations.
\end{itemize}

\subsubsection{Operational Improvements}
\begin{itemize}
    \item \textbf{Automated Model Selection:} Develop frameworks for automated model selection and configuration based on specific deployment environments and operational requirements.
    
    \item \textbf{Failure Pattern Analysis:} Conduct deeper analysis of fail-slow patterns to identify potential correlations with specific workload characteristics or environmental factors.
    
    \item \textbf{Cross-Platform Validation:} Extend the evaluation to different storage systems and hardware configurations to validate the generalizability of our findings.
\end{itemize}

\subsubsection{Extended Research Directions}
\begin{itemize}
    \item \textbf{Transfer Learning:} Investigate the potential of transfer learning approaches to reduce the training data requirements for new deployment environments.
    
    \item \textbf{Interpretability:} Enhance model interpretability to provide clearer insights into the factors contributing to fail-slow predictions, aiding in system maintenance and optimization.
    
    \item \textbf{Preventive Measures:} Develop predictive maintenance strategies based on model insights to prevent fail-slow conditions before they impact system performance.
\end{itemize}

These future research directions aim to address current limitations while expanding the practical applicability of fail-slow detection in production environments. The continued evolution of storage systems and increasing data volumes make these improvements essential for maintaining reliable and efficient storage operations.

\bibliographystyle{unsrt}  


\end{document}